\documentstyle[psfig,twocolumn,aps]{revtex}

\newcommand{\lsco}       {$\rm La_{2-x} Sr_x Cu O_4$}

\newcommand{\lsno}       {$\rm La_{2-x} Sr_x Ni O_{4}$}

\newcommand{\lsnot}	 {$\rm La_{5/3} Sr_{1/3} Ni O_{4}$} 
\newcommand{\lnsco}      {$\rm La_{1.48} Nd_{0.4} Sr_{0.12} Cu O_4$}
\newcommand{\niot}       {NiO$_2$}
\newcommand{\tsons}      {$T_{\rm so}^{\rm NS}$}
\newcommand{\tson}       {$T_{\rm so}^{\rm NMR}$}

\newcommand{\tco}        {$T_{\rm co}$}
\newcommand{\toi}        {$T_1^{-1}$}
\newcommand{\nq}         {$\nu_{{\scriptscriptstyle Q}}$}
\newcommand{\ib}         {$I_B$}
\newcommand{\tc}         {$T_c$}

\newcommand{\td}         {$T$-dependenc}
\newcommand{\hprp}       {$H_0 \!\! \perp \!\! c$}
\newcommand{\hpar}       {$H_0 \!\! \parallel \!\! c$}
\newcommand{\la}         {$^{139}$La}

\draft
\author{Y.~Yoshinari,$^{1}$ P.~C.~Hammel$^{1}$ and S-W.~Cheong$^{2}$}

\title{Magnetism of Stripe-Ordered La$_{5/3}$Sr$_{1/3}$NiO$_{4}$}

\address{$^{1}$Los Alamos National Laboratory, Los Alamos, New Mexico 87545}
\address{$^{2}$Lucent Technologies, Bell Laboratories, Murray Hill, New Jersey 07974\vspace{-3mm}}

\author{\small(Version date:  February 4, 1999)}

\address{
\parbox{14cm}{\bigskip\rm\small
$^{139}$La nuclear magnetic resonance studies reveal markedly different
magnetic properties of the two sites created by the charged domain wall
formation in La$_{5/3}$Sr$_{1/3}$NiO$_{4}$. 
NMR is slow compared to neutron scattering;
we observe a 30 K suppression in magnetic ordering temperature 
indicating glassy behavior.
Applied magnetic field reorients the in-plane ordered moments with
respect to the lattice, but the relative orientation of the spins
amongst themselves is stiff and broadly distributed. 
}}

\begin{document}
\maketitle

\thispagestyle{myheadings}
\markright{\small LA-UR-97-5150 \hspace{28mm} Accepted for publication in {\it Physical Review Letters}}


Since the discovery of 
high temperature superconductivity in the cuprates, 
the behavior of holes added to a strongly correlated two 
dimensional (2D) antiferromagnet has been a subject of intense interest.
One important aspect of this system is its tendency toward  
inhomogeneous charge distribution\cite{phasesep93}.   
In particular, segregation of doped holes into periodic arrays 
of charged stripes separating hole-free domains 
has been predicted\cite{theory}.    
Stripe ordering has been observed in doped La$_{2}$NiO$_{4}$
\cite{Hayden,swc:lsnochen,Tranquada,swc:lee} 
which remains semiconducting up to very high Sr 
content\cite{nickelateconductivity}. 
The recent observation of similar elastic superlattice peaks 
in the isostructural\cite{nickelatestructure,nickelateAFneutron}  
high-\tc\ superconductor \lnsco\cite{tranq:nature} 
indicates the existence of similar charge ordered structures, 
and suggests these structures may be relevant to 
cuprate superconductivity\cite{ek:prb97,tranq:prl97,z:cuprateSC,neto:prl98}.  
Similarities between these elastic superlattice peaks
and the incommensurate peaks observed in 
inelastic neutron studies of \lsco\cite{cheong:nslsco} have 
been noted\cite{tranq:physica97}.    
Hence, these incommensurate peaks are being reconsidered as possible 
evidence for the presence of dynamic charged stripes in 
the cuprate\cite{tranq:physica97}.

There is clear evidence for stripe formation in doped 2D AF 
systems\cite{Hayden,swc:lsnochen,Tranquada,swc:lee}, 
but detailed microscopic studies of their low energy dynamic and 
static magnetic properties are lacking.  
A detailed elastic neutron diffraction study of the title
compound\cite{swc:lee} has demonstrated charge ordering at 
$T_{\rm co} \, = \, 240\,$K\cite{swc:lee,ramirez,blum} into domain walls or ``stripes.''
The stripes run along two equivalent diagonal
directions $\vec{e}=(1,1)$ or $(1,\bar{1})$ in the tetragonal unit cell 
with dimensions $a_{t}\times a_{t}$ in the basal plane 
where $a_t$ is the
is the lattice parameter corresponding to the Ni-Ni spacing.
The stripe period perpendicular to the stripes is 
$3a_t/\sqrt{2}$.
At $T_{\rm so}^{\rm NS} = 190\,$K, spin superlattice peaks appear
indicating ordering of the spins between the stripes. 
Three temperature regions were evident, 
the highest between \tsons\ and \tco\ 
exhibited elastic charge order peaks with weaker intensity 
and significantly reduced correlation 
lengths compared to lower temperatures. 
The authors proposed the existence of a ``stripe glass'' in this 
temperature regime, but were unable to study the 
very important issue of orientational order due to the 
dominant affect of short stripes.  
Below \tsons\ the charge stripe order was found to improve. 

In this Letter we report a single crystal $^{139}$La NMR study of 
the two magnetically distinct sites observed below \tco, 
the first located in the domain walls, 
and the second in the hole-free domains. 
Although the two regions are spatially proximate and strongly interacting, 
their static and dynamic magnetic properties are quite different. 
The NMR provides specific evidence in support of the stripe-glass
hypothesis and reveals pronounced and unusual spin disorder arising from
stripe defects. 
A hallmark of glassy systems is sensitivity of the transition
temperature to measurement time scale; we find the onset of spin
ordering at \tson$= 160\,$K in the NMR measurement is 30 K lower than
\tsons\ as a consequence of its slower characteristic time scale
($\mu$sec compared to psec for the neutron measurement). 
At low temperature where charge order becomes very good, 
the static spin order exhibits a continuous distribution of 
moment magnitudes and in-plane orientations. 
The \hprp\ lineshape demonstrates that the ordered moments rotate 
in response to the applied field such that the most probable spin
orientation is perpendicular to the applied field.
However the broad continuous distribution reveals ``stiffness'' of the
spin texture: defects in the charge order (stripe ends or bifurcations)
frustrate the spin order causing the orientation of the
in-plane ordered moments to rotate relative to each other. 
Finally, the NMR measurements rule out motion of domain
walls below \tco.

A local probe of electronic properties, 
nuclear magnetic resonance (NMR) is well suited to determining 
the microscopic properties of inhomogeneous structures. 
\la\ NMR is employed because $^{61}$Ni NMR is not readily performed.
In \lsno\ a given $^{139}$La nucleus 
($I=7/2$, gyromagnetic ratio $\gamma= 601.44 \,$Hz/Oe) 
directly probes the magnetic properties of only one 
Ni electronic spin moment.
This is because the transferred hyperfine coupling 
of the \la\ nuclear spin to the single Ni site 
that shares the La-apical O-Ni bond 
is an order of magnitude larger than its coupling to 
any of its four other Ni neighbors.  
In the Ni$^{2+}$ (3d$^{8}$) ionic state, 
the two unpaired spins occupy 
the in-plane 3d$_{x^{2}\text{-}y^{2}}$ and the 
out-of-plane 3d$_{3z^2\text{-}r^2}$ atomic orbitals. 
The latter spin is particularly 
\begin{figure}[tbp]
\psfig{file=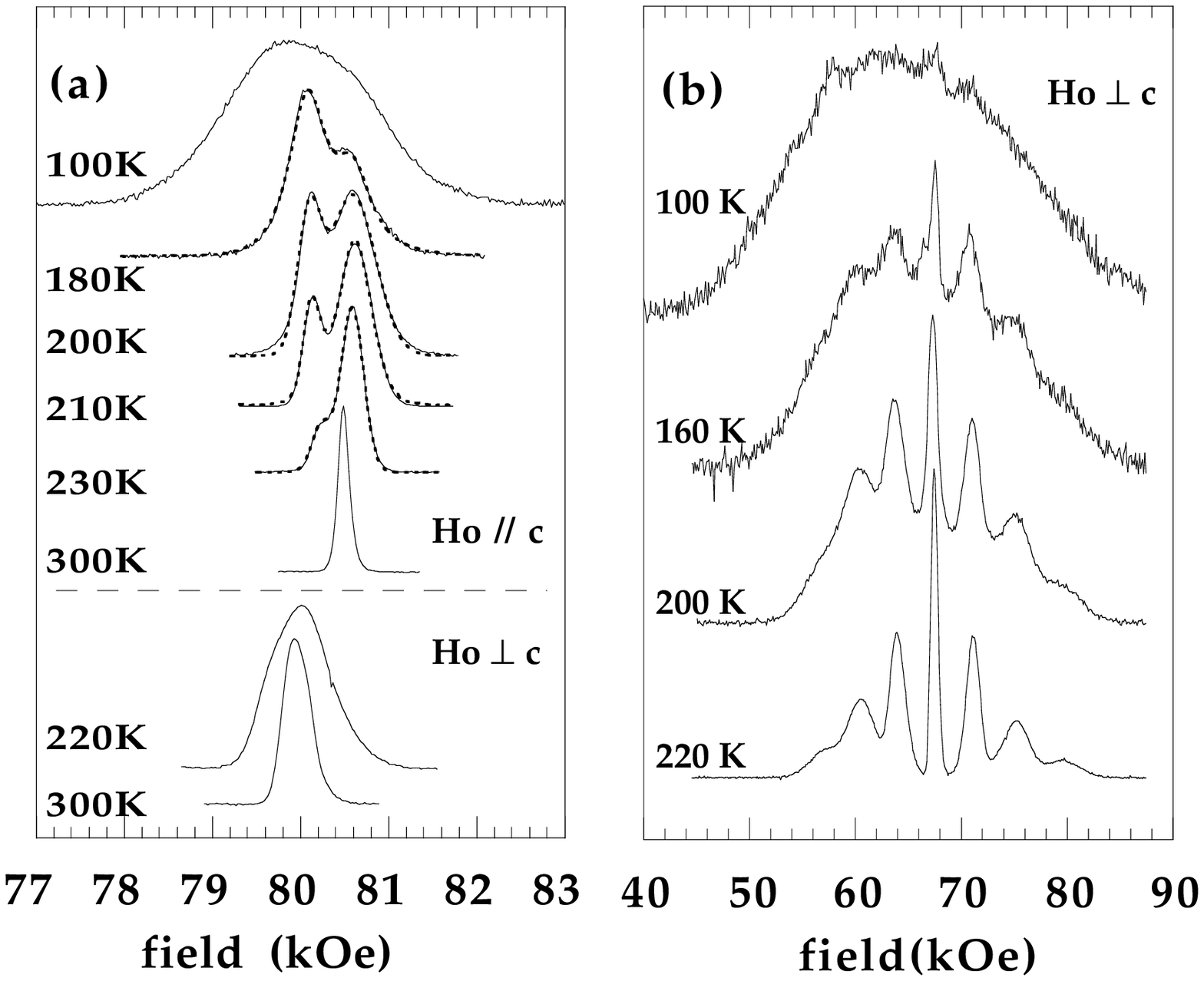,width=3.5in}
\caption{
Quadrupole split $^{139}$La NMR spectra in
La$_{5/3}$Sr$_{1/3}$NiO$_{4}$. 
All spectra were taken using a 
$\frac{\pi}{2}\text{-}30\mu\text{sec-} \pi$ spin-echo sequence.
(a) Central transition ($\text{-}\frac{1}{2}\leftrightarrow \text{+}\frac{1}{2} $)
spectra with \hpar\ (upper),
and \hprp\ (lower), taken at $48.5 \,$MHz.
Dotted lines are fits (see text).
(b) Full spectra including quadrupole satellites taken at 41.0$\,$MHz
with \hprp.}
\label{fig:spc}
\end{figure}
\noindent
strongly coupled 
to the La nuclear moment through an orbital hybridization 
mediated by the apical oxygen. 
The magnitude of the hyperfine field at the La site due to the Ni moment
was determined from a plot of our data for the Knight shift $K$ in the
paramagnetic state vs.\ the 
magnetic susceptibility $\chi$ with temperature as an 
implicit parameter\cite{Kkai} (see also \cite{wada}). 
The Knight shift $K = A \chi $ where $A$ is the hyperfine coupling 
and $\chi$ is the local electronic susceptibility.  
The hyperfine coupling to a moment oriented 
parallel (perpendicular) to the $c$-axis 
is $A_c = 23 \pm 5 \mbox{ kOe/} \mu_B$ 
($A_{ab} = 20{\scriptstyle \pm}2\,$kOe/$\mu _{B}$).  
These fields are one order of magnitude larger than the classical dipole fields of
the same magnetic moments\cite{dipolefieldCu},
so the transferred coupling to the single Ni dominates. 

The 700$\,$mg single crystal of La$_{5/3}$Sr$_{1/3}$NiO$_{4}$ used in this study 
was prepared by the floating zone method\cite{swc:lee}.
The spectrum taken at $220 \,$K shown in Fig.~\ref{fig:spc}(b) 
shows the quadrupole split $^{139}$La NMR field swept spectrum\cite{NMRtext}.
The central 
$\left( -\frac{1}{2}\leftrightarrow {\small+}
\frac{1}{2}\right)$ nuclear Zeeman transition 
is least affected by the distribution of
nuclear quadrupole frequencies \nq\ and so 
produces the most narrow and intense peak\cite{commentnuQ}. 
The Knight shift $K$ and spin-lattice relaxation time $T_{1}$ were measured 
on this peak.
$T_1$ was measured by monitoring the spin-echo intensity
as a function of time after a single saturation pulse was applied.
All relaxation curves were best fit by the ($I=7/2$) theoretical expression for
magnetic coupling as opposed to quadrupolar coupling\cite{recoverycurve}.
For poorly resolved peaks at lower $T$, the measurements
were performed at the value of applied field where the distinct peaks appeared 
at the lowest $T$ for which they were clearly resolved.

Fig.~\ref{fig:spc}(a) 
shows $^{139}$La NMR central transition spectra.
A low field peak 
(the B-peak) emerges from the main A-peak 
below \tco.
The distinct peaks are expected 
since the magnetic environment of Ni sites in or near 
the charged domain walls is very different from that 
of the intervening hole-free regions.
As we will show below, an internal field due to static 
ordered moments contributes to the shift of the 
B-peak 
at low temperature demonstrating the assignment 
of the B-peak 
to the AF-ordered, hole-free regions between domain walls;
the A-peak corresponds to domain wall sites.  

\begin{figure}[tbp]
\vspace{-6mm}
\parbox{1mm}{\rule{0mm}{5mm}}
\parbox{5cm}{\psfig{file=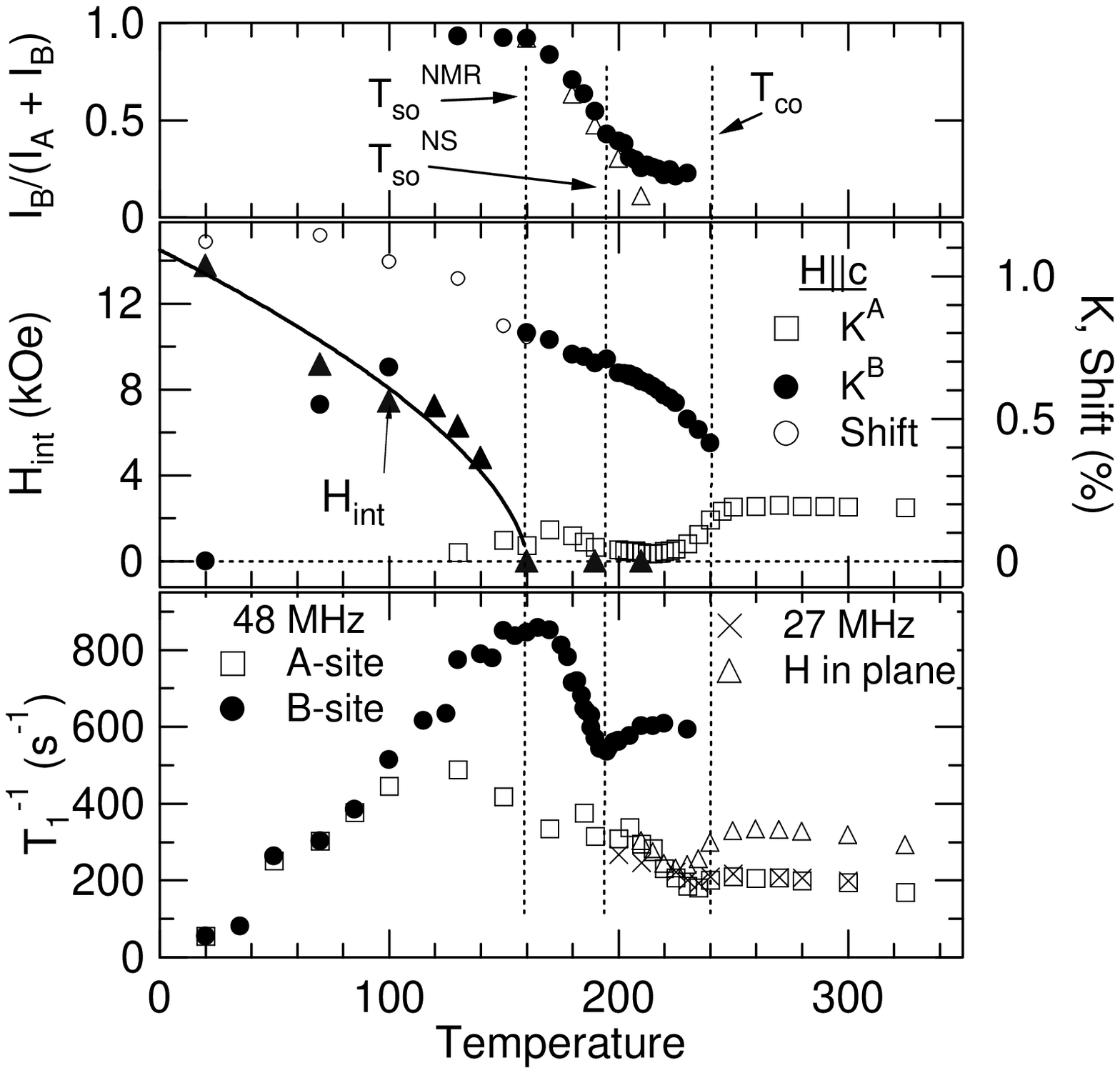,width=3.2in}}
\vspace{1.5mm}\caption{
(a) The \td e of \ib, the B-peak fraction of the
total intensity with \hpar\ ($\bullet$), 
and of the broad peak fraction
for \hprp\ (${\scriptstyle \triangle}$).
(b) The \td ies of the total $^{139}$La shift below 
\tson\ for the B-peak ($\circ$) 
and the Knight shifts $K$ for \hpar\ 
(plotted against the right-hand axis): 
A-peak ($\Box$) and B-peak ($\bullet$). 
The open circles indicate the total shift below \tson.  
$H_{\text{int}}$ is plotted against the left-hand axis.
The solid line is a guide to the eye.
(c) \td ies of the $^{139}$La \toi  for \hpar: 
A- ($\Box$) and B-peak ($\bullet$); 
and for \hprp\ (${\scriptstyle \triangle}$); 
all measured at 48.5$\,$MHz.
\toi\ for the A-peak with \hpar\ at 27$\,$MHz ($\times$) is also shown. }
\label{KandT1}
\end{figure}

With cooling, intensity is smoothly shifted from the 
A- to the B-peak as shown in Fig.~\ref{KandT1}(a) 
and the spectra broaden increasingly rapidly;  
below $\sim $100$\,$K the two lines merge into a
broad Gaussian peak\cite{motionpossibility}.
The \td e of $I_{B}$, the intensity of the 
B-peak relative to the total central transition intensity, 
is displayed in Fig.~\ref{KandT1}(a)\cite{intensityestimation}.
The nuclear spin-spin relaxation time $T_{2}$,
was found to be quite different for the two peaks;
$I_B$ is corrected for $T_{2}$. 
Thus, $I_B$ quantitatively represents the number of nuclear sites 
in the ``B-environment.''
For $210 \text{ K} < T < T_{\rm co}$, 
$I_B \approx \frac{1}{4}$, less than either $\frac{1}{3}$ 
or $\frac{2}{3}$, 
values expected for narrow, sharply defined ordered stripes 
at $\frac{1}{3}$ hole doping. 
With decreasing $T$, \ib\ increases steadily saturating 
at $\approx 0.93$ around 
160 K, indicating that essentially all sites 
are situated in the B-environment.

The internal field due to static spin ordering is apparent in the field
dependence of the shift of the B-peak in the \hpar\ central transition
spectra. 
Because the ordered 
\begin{figure}
\parbox{3mm}{\rule{0mm}{40mm}}
\parbox{0.85\textwidth}{
\psfig{file=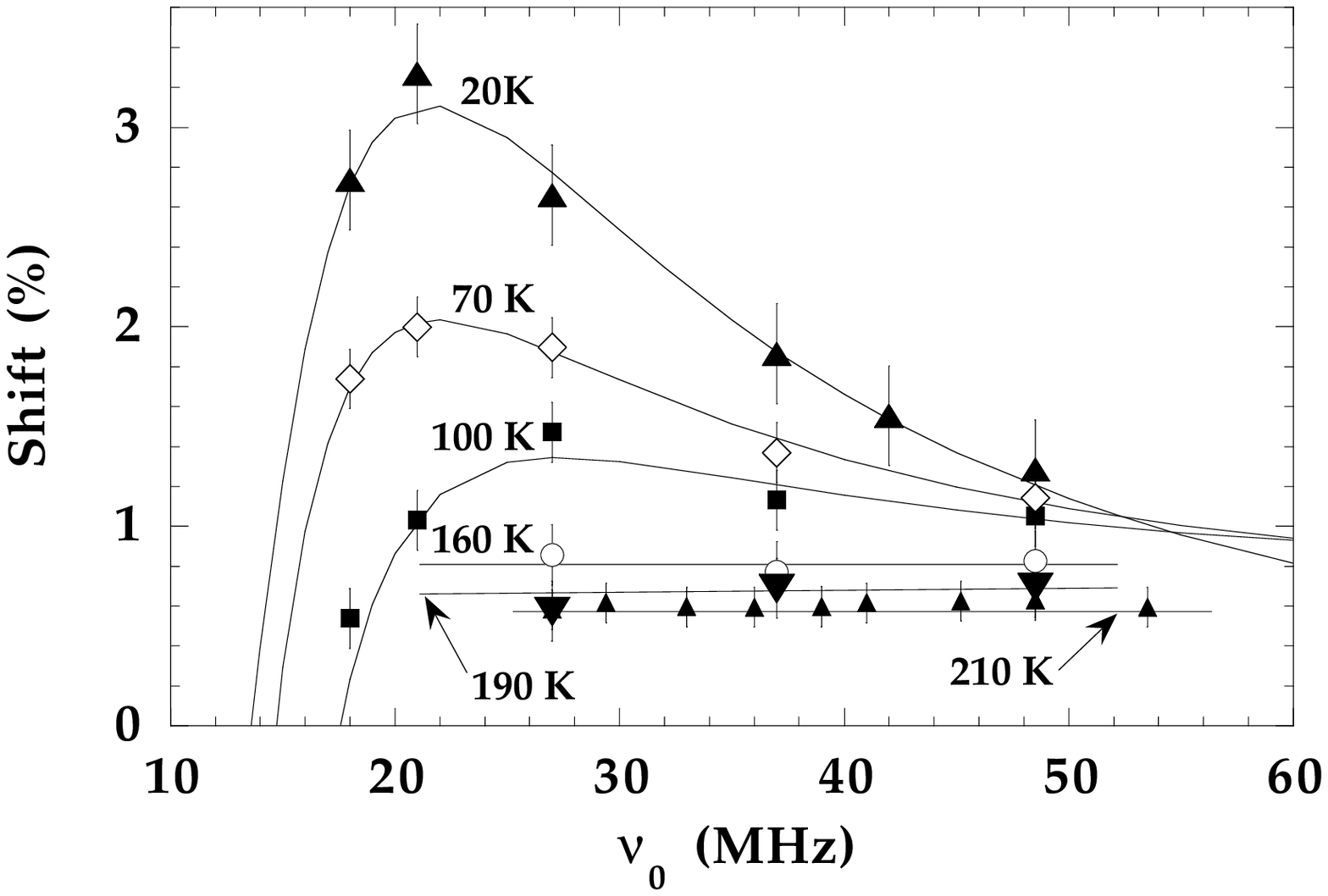,width=2.8in}}
\caption{
The  $\nu_{0}$ dependence of the B-peak 
shift is shown for \hpar\ for various $T$.
The lines show fits to a model incorporating quadrupole 
interactions and an internal field 
lying in the planes ($\perp c$); 
see Eq.~\protect\ref{eq:flddep} and following text.
}
\label{internalfield}
\end{figure}
\noindent
moments lie in the \niot\ planes\cite{Tranquada},
their internal field adds to the applied field in quadrature $\left(
H^2_{\text{eff}} = H^2_0 + H^2_{\text{int}} \right)$. Thus, they do not
produce the large shift one would expect based on the value of $A_{ab} =
20\,$kOe/$\mu _B$. 
The $\nu_{0}(=\gamma H_0)$ dependence of the shift is shown 
at several temperatures for \hpar\ in 
Fig.~\ref{internalfield}.
Understanding the field dependence requires taking account 
of the quadrupole interaction 
which produces a large negative shift at low applied 
fields $\left( \propto - \nu^2_{\scriptscriptstyle {Q}}/\nu^2_0 \right)$. 
Exact diagonalization of 
the Hamiltonian  
\begin{equation}
{{\cal H} \over h} 
 = (1+K) \nu_{0} I_{z}
+\frac {\nu_{\scriptscriptstyle{Q}}} {2} \left[ I_{z}^{2} - \frac {I(I+1)} {3} \right]
+ \nu_{\rm int} I_{x} 
\label{eq:flddep}
\end{equation}
(with $z\parallel c$, and assuming that the moment 
lies in the plane) provides excellent fits (solid lines) 
to the measured field-dependent shifts.
The quadrupole frequency $\nu_{\scriptscriptstyle{Q}}$ was monitored 
and varies by less than 10\%; 
this has negligible affect on the determination of $\nu_{\rm int}$.
The \td e of the
internal magnetic field ($H_{\rm int} \, = \, \nu_{\rm int} / \gamma$) 
deduced from these fits is shown in Fig.~\ref{KandT1}(b). 
We observe no internal field in the 30 K region between \tsons\ where
superlattice peaks are first observed and $T_{\rm so}^{\rm NMR}$ below
which an internal field is observed. 
At 20$\,$K $H_{\text{int}} \simeq 14 \,$kOe at the La site, implying a 
Ni spin polarization 
\( \left< S \right> \mu_B = H_{\text {int}}/A_{ab} \approx 0.7 \mu_B \).
The growth of $H_{\text {int}}$ below \tson\ is unusually slow;
this indicates suppression of the ordered moment by strong magnetic fluctuations. 

At higher temperatures, in the absence of an internal field, 
the Knight shift $K$ is given by Eq.~\ref{eq:flddep} with 
$\nu_ {\text {int}}=0$. 
Above \tson\ the \hpar\ shifts for the A- and B-sites are essentially
independent of $\nu_{0}$, 
demonstrating that they are both exclusively magnetic in origin.
The \td e of $K$ is shown in Fig.~\ref{KandT1}(b). 
$K^{A}$ for the A-sites decreases below \tco, while $K^{B}$ for the
B-sites increases strongly reaching $\sim 0.8 \%$ at \tson. 
This large increase in $K^B$ cannot be attributed to a variation in the
hyperfine coupling since a shift in $A$ would have a much more dramatic
effect on \toi\ ($\propto A^2$) which is not observed; 
furthermore, the variation of the lattice parameter is smooth across
\tco\ and elsewhere\cite{latticeconst}. 
Therefore, the large magnetic shift of the B-sites indicates an
anomalously large susceptibility in the magnetic domains between
stripes.
When the contribution of the static field to the shift is taken into 
account we find that $K^B$ decreases significantly below \tson,
indicating a hardening of the ordered spins to a field applied 
perpendicular to the orientation of the ordered moments. 

Substantial disordering of the in-plane orientation of the static moments
is evident from the broad \hprp\ line [Fig.~\ref{fig:spc}(b)].
The internal field due to the antiferromagnetically ordered moment 
can vary from parallel to anti-parallel orientation relative 
to the field applied in the plane, 
causing shifts comparable to the internal field.   
The half-width of the \hprp\ line is consistent with the  $\sim 12\,$kOe 
ordered moment detected in \hpar\ measurements. 
An orientationally ordered moment would produce discrete lines 
with shifts determined by the orientation 
of the ordered moment relative to the applied field; 
on the other hand, a completely random distribution of in-plane
orientations would generate a two-dimensional
powder pattern spectrum with singularities at its extrema
[distribution $P(\Delta \nu)$  of frequency shifts
\( \sim 1 / \sqrt{(\gamma H_{\rm int})^2 -(\Delta \nu)^2} \)]. 
Neither is seen; instead, the broad spectrum peaked at very small shift
indicates that the most probable in-plane spin orientation is
perpendicular to the applied field. 
This is the usual response of an ordered antiferromagnet since this
allows a slight canting of the moments in response to the field. 
However, the broad width demonstrates a broad distribution of
orientations around the perpendicular orientation.
Thus, an applied field easily rotates the moments in the plane with
respect to the lattice (and, hence, the ordered stripe structure), but
the broad distribution of orientations of the moments relative to one
another is unchanged by applied field. 
This is not due to a fortuitous choice of field
orientation:  the twinning of the stripe orientation would produce two
populations of ordered moments rotated 90${^\circ}$ with respect to each
other, but the spectrum shows no evidence of a second, rotated spin
population.
This in-plane ``twisting'' of the spin texture by stripe defects
presumably occurs over length scales comparable to the spin
coherence length observed with neutrons ($\sim 100\,$\AA\cite{swc:lee}). 

The temperature dependencies of \toi\ for the two environments 
[shown in Fig.~\ref{KandT1}(c)] are very different;
two transitions occur in the hole-free domains (B-line) 
with little effect on spin dynamics in the domain walls (A-line).
At all temperatures signal recoveries demonstrate a single $T_1$ and 
no signal intensity is lost.
Below \tco, \toi\ at the A-sites is monotonic in temperature down to 100
K. In contrast, \toi\ at the B-sites shows a distinct and unusual {\it
negative-going} cusp at \tsons\ that clearly signals the transition at
which spin-superlattice peaks appear in the neutron
measurements\cite{swc:lee}.
This demonstrates the suppression of spin ordering in NMR is not a
consequence of sample variability since the transition at \tsons\ is
clearly observed in our NMR measurements. 
A typical magnetic ordering transition would generate a {\it positive}
cusp or divergence in \toi\ at the ordering temperature 
(see e.g., Ref.\ \onlinecite{Suh}). 
The increase of \toi\ toward a strong peak at lower temperature $ \sim
160$ K indicates a rapid slowing of collective spin fluctuations
indicative of long-range, 3D magnetic order at this temperature. 
The appearance of the internal field at the same temperature would
naturally be interpreted as a magnetic transition, were it not that the
spin-superlattice peak in neutron scattering occurs at the higher
temperature. 

This contrast in apparent ordering temperatures suggests sensitivity to
measurement time scale. 
Rotation of the ordered moment (which is slow compared to the inverse of
the energy resolution of the neutron measurement) could motionally
average, and thus, reduce the measured $H_{\text{int}}$. 
\tson\ would then correspond to the temperature at which this motion
becomes slow on the NMR time scale ($\sim$ microseconds corresponding
to $< 0.1 \mu$eV). 
Our results rule out motion of domain walls below \tco\ even on the NMR
time scale, as we observe distinct relaxation rates for the two sites
immediately below \tco\cite{motionpossibility,yy:tbp}. 
A second possibility is that the magnetic order is sufficiently
sensitive to the applied magnetic field that it is suppressed between 
160 and 195K. 

In summary, the onset of magnetic order in charge-stripe ordered \lsnot\
is suppressed by 30 K when detected on the $\mu$sec time scale of NMR
compared to psec for neutron diffraction; 
this provides strong and specific evidence in support of the 
glassy character of the stripe order.  
The interplay between spin and charge order is evident in
the very broad NMR line found at low temperature where the stripes are
relatively well ordered.
This probably arises from the strong spin disordering
affects of defects (stripe ends and bifurcations) 
in the charge stripe pattern.  
These defects induce a distribution of {\it relative} spin orientations.
However, the collection of spins as a whole is easily reoriented with
respect to the lattice by an applied magnetic field.
Finally, below the charge ordering temperature, the ordered charge
stripes themselves are static on the NMR time scale. 

We gratefully acknowledge stimulating discussions with A. R. Bishop, 
D. E. MacLaughlin, A. J. Millis, Byoung Jin Suh and Z. G. Yu. 
One of us (PCH) appreciates the contributions of J. Zaanen 
to our understanding of these results.  
Work at Los Alamos was supported by the Department of Energy, Office of 
Basic Energy Sciences.



\vspace{-2mm}
\begin{references}
\vspace{-10mm}
\bibitem{phasesep93}
{\em Proceedings of the Workshop on Phase Separation in Cuprate
  Superconductors}, edited by K.~A. M{\"u}ller and G. Benedek (World
  Scientific, Singapore, 1993).

\bibitem{theory}
J.~Zaanen and O.~Gunnarsson, Phys.\ Rev.\ B \textbf{40}, 7391 (1989);
D.~Poilblanc and T.~M.~Rice, Phys.\ Rev.\ B \textbf{39}, 9749 (1989);
H.~J.~Schulz, J. Phys. (Paris), \textbf{50}, 2833 (1989); 
V.~J.~Emery, S.~A.~Kivelson and H.~Q.~Lin, Phys.\ Rev.\ Lett.\ \textbf{64}, 475 (1990); 
H.~E.~Vierti\"{o} and T.~M.~Rice, J. Phys. Condens. Matter \textbf{6}, 7091 (1994);
J.~Zaanen and P.~B.~Littlewood, Phys.\ Rev.\ B \textbf{50}, 7222 (1994); and
S.~R.~White and D.~J.~Scalapino, Phys.\ Rev.\ Lett.\ {\bf 80}, 1272 (1998).
 
\bibitem{Hayden}
S.~M.~Hayden \textit{et al.}, Phys.\ Rev.\ Lett.\ \textbf{68},
1061 (1992).

\bibitem{swc:lsnochen}
C.~H.~Chen, S-W.~Cheong and A.~S.~Cooper, Phys.\ Rev.\ Lett.\ \textbf{71},
2461 (1993).

\bibitem{Tranquada}
J.~M.~Tranquada \textit{et al.}, Phys.\ Rev.\ Lett.\ \textbf{73},
1003 (1994); 
J.~M.~Tranquada, D.~J.~Buttrey and V. Sachan, Phys.\ Rev.\ B \textbf{54},
12318 (1996).

\bibitem{swc:lee}
S.-H.~Lee and S-W.~Cheong, Phys.\ Rev.\ Lett.\ \textbf{79}, 2514 (1997).

\bibitem{nickelateconductivity}
S-W.~Cheong \textit{et al.}, Phys.\ Rev.\ B \textbf{49}, 7088 (1994).

\bibitem{nickelatestructure}
J.~D.~Jorgenson \textit{et al.}, Phys.\ Rev.\ B \textbf{40}, 2187 (1989).

\bibitem{nickelateAFneutron}
G.~Aeppli and D.~J.~Buttrey, Phys.\ Rev.\ Lett.\ \textbf{61}, 203 (1988);
K.~Nakajima \textit{et al.}, Z. Phys.\ B \textbf{96}, 479 (1995).

\bibitem{tranq:nature} J.~M.~Tranquada \textit{et al.},
Nature, \textbf{375}, 561 (1995);


\bibitem{ek:prb97}
V.~J.~Emery and S.~A.~Kivelson, O.~ Zachar, Phys.\ Rev.\ B {\bf 56}, 6120 (1997). 

\bibitem{tranq:prl97}
J.~M.~Tranquada \textit{et al.}, Phys.\ Rev.\ Lett.\ \textbf{78},
338 (1997).

\bibitem{z:cuprateSC}J. Zaanen, M. L. Horbach and W. van Saarloos, 
Phys.\ Rev.\ B {\bf 53}, 8671 (1996). 

\bibitem{neto:prl98} 
A.~H.~Castro Neto and F. Guinea, Phys.\ Rev.\ Lett.\ {\bf 80}, 4040 (1998). 

\bibitem{cheong:nslsco}
S-W. Cheong \textit{et al.}, 
Phys.\ Rev.\ Lett.\ {\bf 67}, 1791 (1991).

\bibitem{tranq:physica97} J. M. Tranquada, Physica C {\bf 282}, 166 (1997).

\bibitem{ramirez}
A.~P.~Ramirez \textit{et al.}, Phys.\ Rev.\ Lett.\ \textbf{76}, 447 (1996).

\bibitem{blum}
G. Blumberg, M. V. Klein and S-W. Cheong, 
Phys.\ Rev.\ Lett.\ \textbf{80}, 564 (1998).


\bibitem{Kkai}
A.~M.~Clogston and V.~Jaccarino, Phys.\ Rev.\ {\bf 121}, 1357 (1961)

\bibitem{wada} S. Wada, {\it et al.,} J. Phys.\ Soc.\ Jpn.\ {\bf 58}, 2658 (1989).

\bibitem{dipolefieldCu} Cu NQR measurements in La$_{2}$CuO$_{4}$ indicate  
the transferred hyperfine coupling to the four nearest Ni 
sites via the La(6s) orbital will be of the same order as the dipole field; 
H.~Nishihara \textit{et al.}, J. Phys.\ Soc.\ Jpn.\ {\bf 56}, 4559 (1987).

\bibitem{NMRtext}
C.~P.~Slichter, \emph{Principles of Magnetic Resonance}, 3rd ed.
(Springer-Verlag, New York, 1990);
A.~Abragam {\em The Principles of Nuclear Magnetism},
(Oxford University Press, London, 1961).

\bibitem{commentnuQ}
$\nu_{{\scriptscriptstyle Q}}$ was found to be significantly
distributed at all temperatures 
(at $300\,$K $\nu _{{\scriptscriptstyle Q}}\simeq 4.55\,$MHz with width
$1.7\,$MHz).
Essentially no change in $\nu _{{\scriptscriptstyle Q}}$
was observed at \tco.

\bibitem{recoverycurve} A.~Narath, Phys.\ Rev.\ {\bf 162}, 320 (1967);
D.~E.~MacLaughlin \textit{et al.}, Phys.\ Rev.\ B \textbf{4}, 60 (1971).

\bibitem{motionpossibility}
We carefully explored the possibility of charge stripe 
motion\protect\cite{yy:tbp}.  
The absence of frequency dependence in 
\toi\ and the sharp difference in \toi\ at the 
two sites demonstrates that the charge stripes 
are static on the time scale $10^{-3}\,$s.

\bibitem{intensityestimation}
Between 190$\,$K and \tco\ intensity was estimated by fitting each peak
with a single Gaussian; 
below 190$\,$K a product of a Gaussian and Lorentzian was used.
By varying $\tau$ a nearly pure A-peak spectrum was obtained which
confirmed that this deconvolution method nicely reproduces the observed
lineshape.
The single exponential spin-echo decay over more than a decade 
verifies the single-component nature of each peak\protect\cite{yy:tbp}. 

\bibitem{latticeconst} S.~H.~Han \textit{et al.},
Phys.\ Rev.\ B \textbf{52}, 1347 (1995).

\bibitem{Suh}
B.~J.~Suh \textit{et al.},
Phys.\ Rev.\ Lett.\ \textbf{75}, 2212 (1995).

\bibitem{yy:tbp} Y. Yoshinari \textit{et al.}, to be published.

\end{references}
\end{document}